\documentclass[prb,preprint, showpacs,aps]{revtex4}
\usepackage{latexsym}
\usepackage{graphicx}
\usepackage[activeacute,english]{babel}

\newcommand{\mean}[1]{\left \langle #1 \right \rangle}

\begin{document}
   
\title{Optimal protocols for minimal work processes in underdamped stochastic thermodynamics}
  
\author{Alex Gomez-Marin$^{1,2}$, Tim Schmiedl$^{2}$, and Udo Seifert$^{2}$}

\affiliation{$^1$ Facultat de Fisica, Universitat de Barcelona, Diagonal 647, 08028 Barcelona, Spain
\\
$^2$ II. Institut f\"{u}r Theoretische Physik, Universit\"{a}t Stuttgart, 70550 Stuttgart, Germany
} 

\date{\today}

\begin{abstract}
For systems in an externally controllable time-dependent potential, the optimal protocol minimizes the mean work spent in a finite-time transition between two given equilibrium states. For overdamped dynamics which ignores inertia effects, the optimal protocol has been found to involve jumps of the control parameter at the beginning and end of the process. Including the inertia term, we show that this feature not only persists but that even delta peak-like changes of the control parameter at both boundaries make the process optimal. These results are obtained by analyzing two simple paradigmatic cases: First, a Brownian particle dragged by a harmonic optical trap through a viscous fluid and, second, a Brownian particle subject to an optical trap with time-dependent stiffness. These insights could be used to improve free energy calculations via either thermodynamic integration or ``fast growth'' methods using Jarzynski's equality.

\end{abstract}


\maketitle
\section{Introduction}

The free energy difference $\Delta F$ between two equilibrium states is an important quantity in isothermal statistical mechanics. Strategies to extract $\Delta F$ from experiments or computer simulations are traditionally based on either thermodynamic integration  or thermodynamic perturbation \cite{zwan54} which use one infinitesimally slow transition or many infinitesimally fast transitions, respectively, between the two equilibrium states. A decade ago, Jarzynski proposed the remarkable relation 
\begin{equation}
e^{-\Delta F /  T} = \mean{e^{- W / T}}
\end{equation}
which interpolates between these extreme cases using nonequilibrium work values $W$ obtained from trajectories of finite time transitions between the equilibrium states at temperature $T$ (with Boltzmann's constant $k_B = 1$ throughout the paper) \cite{jarz97, jarz97a}. This exact relation which holds for any time-dependent driving described by an external control parameter $\lambda(t)$ has been extended to various fluctuation theorems \cite{croo99, croo00, humm01,hata01,seif05a}. Although these (necessarily irreversible) finite time transitions occur in nonequilibrium, the equilibrium quantity $\Delta F$ can be inferred from a sufficient number of trajectories either from computer simulations \cite{hend01, humm01a, shir03, park04, mara06} or real experiments \cite{liph02, coll05}. However, the convergence of the involved exponential average causes problems for far out of equilibirium transitions where the work $W$ is substantially larger than the free energy difference $\Delta F$ \cite{gore03}. In this regime, the exponential average is dominated by low work values which are very rarely sampled \cite{jarz06}. As a remedy, several path sampling techniques biasing the dynamics for low work have been proposed \cite{ytre04, atil04, lech06}. It is, however, still under debate \cite{ober05, ytre06, lech07}  for which systems fast growth techniques are superior to refined ``conventional'' approaches such as umbrella sampling \cite{torr77} or flat histogram methods \cite{wang01}. 
Though valuable for computer simulations, it is hard to imagine how to bias dynamics in real experiments, where, however, apparatus drift may prevent long measurements necessary for thermodynamic integration \cite{coll05, mara07} and thus render fast growth methods competitive. 

Both for thermodynamic integration and ``fast growth'' methods employing Jarzynski's equality (or some variant), efficiency gains can be achieved by optimizing the driving scheme $\lambda(t)$. For thermodynamic integration, where the work $W \geq \Delta F$ is taken as an estimator for $\Delta F$, it is obvious that a minimal work gives the best result \cite{koni05}. In the case of fast growth methods, the statistics for free energy estimates quite generally also improves with smaller mean work \cite{atil04, jarz06}. Fast growth simulations even allow to combine data from different driving schemes in a straightforward way \cite{minh06}.

The minimization of the work spent in a finite time process can, however, also be seen in the context of minimal energy dissipation. On a macroscopic scale, the optimization of the work (or power) exerted in a macroscopic cyclic process has been discussed for quite a while \cite{curz75, band82, andr84, hoff03}. On a microscopic scale, fluctuations will also affect optimal cyclic processes \cite{schm08} which may become relevant for constructing optimal nano machines.

For overdamped Langevin dynamics, the optimal protocol leading to a minimal mean work in a finite time $t$ has been calculated analytically for harmonic potentials \cite{schm07}. Surprisingly, the optimal protocol shows jumps at the beginning and end of the finite time transition. Since most molecular dynamics (MD) simulations of the dynamics of biomolecules are on time-scales where inertia plays an important role (see \cite{soto07} for a review on steered MD), it is an interesting question how these results transfer to underdamped dynamics. In particular, it is important to know whether the jumps are a result of having neglected inertia.

In this paper, we investigate the role of inertia for two previously introduced paradigmatic processes. In Sect. \ref{case1}, we calculate optimal driving schemes for an underdamped Brownian particle dragged through a viscous fluid by harmonic optical tweezers. In Sect. \ref{case2}, we study an underdamped Brownian particle subject to an optical trap with time-dependent stiffness. In both cases, we compare our findings with the corresponding results in the overdamped limit \cite{schm07}. We find that the optimal protocol still involves jumps. Even more surprisingly, the optimal protocol includes delta peaks at the beginning and end of the process.

\section{Case Study I: The moving trap}
\label{case1}

\subsection{Optimal protocol}

We consider a Brownian particle of mass $m$
dragged through a viscous fluid with friction coefficient $\gamma$ by a harmonic potential 
\begin{equation}
V(x,t)=\frac{k}{2}(x-\lambda(t))^2,
\end{equation}
where $k$ is the (constant) trap stiffness. The focus of the optical trap $\lambda(t)$ is changed time-dependently from an initial position $\lambda_i = 0$ to a final position $\lambda_f$.
Including inertial effects, the Langevin equation reads
\begin{equation} \label{langm}
m\ddot{x}=-\gamma \dot{x} -k(x-\lambda(t)) + \eta(t),
\end{equation}
where thermal fluctuations are modeled by Gaussian white noise 
\begin{equation}
\langle \eta(t) \eta(t') \rangle = {2 T} {\gamma} \delta(t-t').
\end{equation}
The mean work spent in the process of total duration $t_f$ is given by 
\begin{equation}
 W  \equiv \int_{0}^{t_f}dt  \left\langle \frac{\partial V(x,t)}{\partial t}  \right\rangle,
\end{equation}
where the average $\langle \dots \rangle$ is over the intitial thermal distribution and over the noise history.
In the present case the total (mean) work reduces to
\begin{equation} \label{work0}
 W  = \int_{0}^{t_f}dt k\dot{\lambda}(\lambda - u ) = k \int_0^{tf}  \lambda \dot u + \frac k 2 \lambda_f^2 - k\left [ \lambda u \right]_0^{t_f} ,  
\end{equation}
where, for simplicity in the notation, we have defined the mean position of the particle as $u\equiv \langle x \rangle$. 
This quantity $u(t)$ depends on the whole history of $\lambda(t)$ and thus, the  work $W$ is a non-local functional of the protocol $\lambda(t)$. However, in analogy to the overdamped limit \cite{schm07}, we can express the work as a local functional of the mean particle position $u$. By averaging the evolution equation (\ref{langm}) we have
\begin{equation}\label{l1}
\lambda=u +\gamma \dot{u}/k+m\ddot{u}/k,
\label{lam_u}
\end{equation}
which inserted in Eq. (\ref{work0}) leads to
\begin{eqnarray}
 W  =\left[  \frac{m}{2}\dot{u}^2 +\frac{m^2}{2k}\ddot{u}^2 +\frac{m\gamma}{k}\dot{u}\ddot{u} +\frac{\gamma^2}{2k}\dot{u}^2  \right]_{0}^{t_f}
     +\gamma \int_{0}^{t_f}dt \dot{u}^2 .
\end{eqnarray}

The only term remaining in the integral, $\dot{u}^2$, is identical to the one in the overdamped limit, while the boundary terms are different.  In complete analogy to the overdamped case, we now proceed in two steps. First, we calculate the optimal shape $u(t)$ minimizing only the integral given initial values $u(0^+) = 0$ and  $\dot u(0^+) = A$. Note that despite the initial equilibrium value $\dot u(0) = 0$, we are free to choose $\dot u(0^+) = A$ since the necessary ``kink'' in $u(t)$ at $t = 0$ does not contribute to the integral. Similarly, at the end of the protocol (at $t=t_f$) there can be another jump in the velocity. In a second step, we adjust the constant $A$ to yield the minimal total work.
First, from the Euler-Lagrange equation corresponding to the Lagrangian $\dot{u}^2 $ (and subject to the initial conditions just mentioned), we find 
\begin{equation}
u(t)=At
\label{u} 
\end{equation}
for $0< t < t_f$. 
In contrast to the overdamped case, we cannot determine all the boundary terms at $t_f$ from the evolution equation. Thus, $C \equiv \dot{u}(t_f) $ is another free parameter. With $\ddot{u}(t_f)=[k(\lambda_f-At_f)-\gamma C ]/m$, we get the total work as a function of the yet unknown constants $A$ and $C$
\begin{eqnarray}
 W(A,C) =\frac{m}{2}C^2+ \frac{k}{2} (\lambda_f-At_f)^2  +\gamma  \int_{0}^{t_f}dt A^2.
\end{eqnarray}
The work is clearly minimal for $C^*=0$, where the asterisk will denote optimal from now on. The remaining terms then read
\begin{eqnarray}
W(A) = \frac{k}{2} (\lambda_f-At_f)^2  +\gamma t_f A^2 ,
\end{eqnarray}
which, surprisingly, is exactly the same expression that was found in the overdamped limit. Minimizing this expression with respect to $A$ leads to
\begin{eqnarray}
A^*=\frac{\lambda_f}{2\gamma/k +  t_f}
\end{eqnarray}
which yields the work
\begin{eqnarray}\label{W*}
W^*=k\lambda_f^2 \frac{1}{2+ {k t_f} / {\gamma}}.
\end{eqnarray}
Inserting Eq. (\ref{u}) into Eq. (\ref{lam_u}), we find the optimal protocol
\begin{equation}
\lambda^*(t)=\lambda_f \frac{kt/\gamma+1}{kt_f/\gamma+2},  
\label{mov_lam*}
\end{equation}
for $0 < t < t_f$ implying symmetrical jumps 
\begin{equation}
\Delta \lambda \equiv \lambda(0^+)-\lambda_i
=\lambda_f-\lambda(t_f^-)=
\lambda_f \frac{1}{kt_f/\gamma+2}
\end{equation}
at the beginning and at the end of the process.

Superficially, this optimal protocol looks like the expression in the overdamped case \cite{schm07}. There is, however, a subtle difference arising from the presence of inertia terms. 
The optimal protocol forces the mean velocity to
instantly  jump at the beginning of the process 
 from its initial equilibrium value $\dot{u}(0)=0$ to $\dot{u}(0^+)=A^*$. At the end of the protocol, the optimal strategy consists in setting back the mean velocity to zero $\dot{u}(t_f)\equiv C^*=0$. 
Due to the second time derivative in the equation of motion such jumps in the velocity, which require delta functions in the acceleration, imply a delta-type singularity in the protocol.
Specifically, in Eq. (\ref{l1}), the jumps in $\dot u$ imply a $\delta$-function for $\ddot{u}$ and hence a  $\delta$ function in $\lambda(t)$. The optimal protocol [Eq. (\ref{mov_lam*})] thus becomes
\begin{equation}
\lambda^*(t) = \lambda_f \frac{kt/\gamma+1}{kt_f/\gamma+2} +  {\frac{m \lambda_f}{2\gamma +  k t_f}} \left [ \delta(t) -   \delta(t-t_f) \right ],
\end{equation}
as shown in Fig. \ref{fig1}. In the overdamped limit, $m \to 0$, the delta peaks vanish.

\subsection{Physical origin of singularities in the optimal protocol}

The benefit of having jumps in the optimal protocol can be understood intuitively as follows. From the perspective of minimal dissipation, it is obvious that the particle should be dragged at a constant (mean) velocity from the beginning rather than being accelerated during a finite time. This initial jump in the velocity of the particle can only be achieved by a finite initial difference $\lambda(0) - u(0)$, corresponding to a jump in $\lambda$ at $t = 0$. In the present underdamped regime, a velocity jump corresponds to a jump in the (mean) particle momentum which can only be achieved by a delta peak in the force, corresponding to a delta peak in the protocol.

The final jump is harder to grasp intuitively. In fact, it stems from focussing on the minimal work rather than on the minimal (mean) dissipation (or entropy production). If we had searched for the minimal entropy production (as defined in \cite{seif05a}), we would have found an optimal protocol without a final jump. In the present minimization, at the final time $t_f$, the particle is still in non-equilibrium with respect to the final potential $V(x, \lambda(t_f))$. Relaxation to equilibrium leads to further dissipation {\it after} time $t_f$ which has, however, already been paid for by the total work since at constant $\lambda$ no work is exerted anymore. A smaller final particle position $u(t_f)$ leads to a longer relaxation time which can decrease the total dissipation of the combined process (nonequilibrium transition and relaxation).

The final delta peak corresponds to setting the final velocity to zero. This decreases the kinetic energy of the particle and thus is beneficial for a small work. It also explains the surprising fact that, according to Eq. (\ref{W*}), we do not have to pay any extra cost for having inertia. During the initial singularity, the exerted work is stored in the (mean) kinetic energy of the particle. This contribution is fully recovered during the final singularity where the kinetic energy of the particle is set back to the equilibrium value.

\subsection{Comparison to a linear protocol}

Without prior knowledge, one might have expected a continuous linear protocol 
\begin{equation}
\lambda^{\rm lin}(t) \equiv \lambda_f t / t_f 
\end{equation}
to yield the lowest work. In the overdamped limit, the work for a linear protocol was at most $14 \%$ larger than for the optimal protocol. We now check how much smaller the value of the optimal work $W^*$ is compared to a linear protocol if we include inertia. First, we rescale the system in order to compactly write the relevant combination of parameters.
With the rescaled mass $\tilde{m}\equiv mk/\gamma^2$, the energy scale $e\equiv k\lambda_f^2$ and a rescaled time $\tilde{t}\equiv t_fk/\gamma$, the  work can be written as
$W=e\tilde{W}(\tilde{t},\tilde{m})$, with the optimal work
$W^*=e / \left( 2+\tilde{t}\right)$.

Solving the second order differential equation of motion (\ref{lam_u}) using the linear protocol $\lambda^{\rm lin}(t)$, we find the ratio:
\begin{equation}
\frac{W^{\rm{lin}}}{W^*}
 =  \left \lbrace \begin{array}{l} \frac {2+\tilde{t}} {\tilde{t}^2}  \left ( \theta_0+\tilde{t}
 -e^{- \frac {\tilde{t}} {2\tilde{m}}}  \left [ \theta_0 \cosh \left ( \nu \tilde{t} \right ) + \theta_1 \sinh(\nu \tilde{t}) \right ]   \right )  ~~~\tilde m < \frac 1 4 \\ \frac {2+\tilde{t}} {\tilde{t}^2}  \left (
  \theta_0 + \tilde{t} 
  - e^{- \frac {\tilde{t}} {2\tilde{m}}}  \left [ \theta_0 ~\cos \left ( \nu \tilde{t} \right ) ~+~ \theta_1~ \sin(\nu \tilde{t}) \right ]  \right ) ~~~\tilde m > \frac 1 4  \end{array} \right.
\end{equation}
with
\begin{eqnarray}
\nu = \frac {\sqrt{\left | 4\tilde{m} -1 \right | }} {2\tilde{m}} 
\end{eqnarray}
and
\begin{eqnarray}
\theta_0=\tilde{m} - 1
,\;\; \theta_1=\frac{3\tilde{m}-1}{2\tilde{m}\nu} .
\end{eqnarray}
In Fig. \ref{fig2}, we plot the ratio ${W^{\rm{lin}}} / {W^*}$ as a function of rescaled time $\tilde{t}$ and mass $\tilde{m}$. This result shows that the optimal protocol significantly reduces the work spent in the process compared to a linear protocol.

\section{Case Study II: The Stiffening trap}
\label{case2}

In the first case study, only the averaged quantity $u = \mean{x}$ appeared in the  work and thus the same result could have been obtained from a deterministic damped dynamics. We next examine a second case study where fluctuations are important. We consider a Brownian particle of mass $m$ in an optical trap with time dependent stiffness $\lambda(t)$ which is driven from an initial value $\lambda(0) = \lambda_i$ to a final value $\lambda(t_f) = \lambda_f$ in a finite time $t_f$. The time dependent potential
\begin{equation}
V(x,t)=\frac{\lambda(t)}{2}x^2,
\end{equation}
leads to the underdamped Langevin equations
\begin{eqnarray}
\dot{x} & = & p/m \\ \nonumber
\dot{p} & = & - \gamma p/m -\lambda(t)x + \eta(t),
\end{eqnarray}
where $p$ is the momentum of the particle
and the noise $\eta(t)$ has the same properties introduced in the first case study. 
Again our main goal is to find the protocol for which the corresponding total (mean) work
\begin{equation}\label{W1}
W  = \int_{0}^{t_f}dt  \dot{\lambda} \frac{ \langle x^2 \rangle  }{2}
\end{equation}
is minimal.
Note that the mean squared position 
\begin{equation}
w \equiv \langle x^2 \rangle 
\end{equation}
of the particle is non-trivially coupled to the mean squared momentum 
\begin{equation}
z \equiv \langle p^2 \rangle 
\end{equation}
and to the position-momentum correlation
\begin{equation}
y\equiv\langle xp \rangle . 
\end{equation}
Their time evolution is governed by the set of coupled differential equations
\begin{eqnarray} 
\dot{w} & = &2y/m  \label{wdot} , \\
\dot{z} &= &-2\lambda y  -2 \gamma z/m +2\gamma T \label{zdot} , 
\\
\dot{y} & = & z/m-\lambda w - \gamma y/m. \label{ydot}
\end{eqnarray}

Unlike both the moving trap (with and without inertia) and the stiffening trap in the overdamped limit, the present case is much more involved since one cannot eliminate the protocol and write the work as a function of one variable only.
We thus express the work as a time-local functional of $w(t)$ and $z(t)$. Solving Eqs. (\ref{wdot}) and (\ref{ydot}) for $\lambda$ yields 
\begin{eqnarray} \label{Lstiff}
\lambda=\frac{1}{w}[z/m-\gamma\dot{w}/2-m\ddot{w}/2]
\end{eqnarray}
which, inserted in Eq. (\ref{W1}) and after partial integration, leads to
\begin{equation} \label{W20}
W  =
\left[ \frac{\lambda w}{2}+\frac{m\dot{w}^2}{8w} \right]_{0}^{t_f} + \frac{1}{2}\int_{0}^{t_f}dt  \mathcal{L}
\label{W_stiff}
\end{equation}
with the ``Lagrangian''
\begin{equation} \label{Lagrangian}
\mathcal{L}= \frac{ \gamma \dot{w}^2}{2w}-\frac{z\dot{w}}{mw}+
\frac{m\dot{w}^3}{4w^2}.
\end{equation}

We proceed in two steps analogously to the moving trap. We first minimize the integral in Eq. (\ref{W_stiff}) for given initial conditions and then optimize with respect to remaining free parameters.
The integrand $\mathcal{L}$ depends on $w$ (and $\dot{w}$) but also on $z$. The variables $w$ and $z$ are not independent. 
Eliminating $\lambda$ and $y$ from the equations of motion (\ref{wdot}), (\ref{zdot}), and (\ref{ydot}), we find the physical constraint
\begin{equation}
\label{constr}
\mathcal{G}\equiv z\dot{w}-m\gamma\dot{w}^2 /2-m^2\dot{w}\ddot{w}/2-2\gamma Tw +2\gamma wz/m+w\dot{z}=0.
\end{equation}
A detailed analysis of the solution of this optimization problem using Euler-Lagrange equations is given in Appendix \ref{app1}.

The results for both the rescaled optimal protocol $\lambda^* ( t / t_f) / \lambda_i$ and the optimal work $W^* / T$ depend on
the dimensionless quantities 
\begin{equation}
\tilde{t}\equiv t_f \lambda_i /\gamma ,~~ \tilde{\lambda}\equiv \lambda_f /\lambda_i, ~~\tilde{m}\equiv \lambda_i m /\gamma^2.
\end{equation}
An extensive analysis of the optimal protocol as a function of all three parameters is out of scope. Since the overdamped limit ($\tilde m \to 0$) has been discussed previously, we focus on the behaviour as a function of $\tilde m$ for given $\tilde \lambda = 2$, $\tilde t = 1$. Given these parameters, the optimization problem can be solved numerically and the corresponding total work $W$ can be calculated. 
In Fig. \ref{lam*}a, we plot the value of the minimal work $W^*$ (obtained from the optimal protocol) as a function of the rescaled dimensionless mass $\tilde{m}$ and compare it to other benchmark protocols. All work values are bounded from below by the free energy difference $\Delta F = \left ( \ln 2 / 2 \right )   T \simeq 0.35  T$. Quite generally, work values are also bounded from above by the work for an immediate jump $W^{\rm{jp}} \equiv \lim_{t\to 0} W^* =   T / 2$. We study (i) a linear protocol, (ii) a protocol leading to a parabolic mean-squared position
\begin{equation}
w(t) = \frac T {\lambda_i} (1 + c t)^2
\label{w_harm}
\end{equation}
with optimized parameter $c$ and optimized final delta peak, and (iii) a protocol leading to
\begin{equation}
w(t) = 1 + a t^3 \left ( 1 - e^{-1 / \left [ 0.01+(5 t/t_f)^2  \right ] } \right ) + b t^5 + c t^7 + d t^9
\label{w_poly}
\end{equation}
without any discontinuities (except for a final jump) but with free parameters $a, b, c, d$. The work arising from protocol (i) lies significantly above the optimal protocol. Protocol (ii) implies (optimized) jumps and delta peaks at the beginning and end. The work for protocol (ii) and the optimal work almost coincide. The inset shows that the optimal work is in fact slightly smaller than the work obtained for the protocol (ii). The difference to the numerically obtained exact solution $W^*$ decreases for decreasing $\tilde m$ which is consistent with the analytical finding that protocol (ii) is optimal in the overdamped limit. The fact that protocol (ii) which involves optimized singularities but not the optimized shape is so close to the optimal work highlights the importance of jumps for the optimal protocol. Protocol (iii) has no delta peaks and no initial jump but mimics these features approximately since the parameters $a, b, c, d$ have been optimized, see Fig. \ref{lam*}b. These trial protocols show that jumps and delta peak-like singularities can decrease the total work and confirms that our numerical solution of the Euler-Lagrange equations is the solution of the optimization problem. 

Finally, the explicit shape of the optimal protocol $\lambda^*(t)$ can be reconstructed numerically from Eq. (\ref{Lstiff}), see Fig. \ref{lam*}c. It displays initially a delta peak upwards accompanied with a jump $\Delta \lambda$ and, finally, a delta peak downwards together with another jump $\Delta \lambda '$. Such discontinuities in the protocol are a consequence of the discontinuities in $z$, $\dot{w}$ and $\ddot{w}$. The first singularity is ``needed'' to suddenly increase $\langle p^2\rangle$ from its equilibrium value and also to change the derivative of $\langle x^2\rangle$, which is proportional to the correlation $\langle xp \rangle$. Note that both size and direction of the jumps strongly depend on the rescaled mass $\tilde m$ as shown in Fig. \ref{lam*}d. For small $\tilde m$, the protocol jumps upwards (as also observed in the overdamped regime \cite{schm07}) while for large $\tilde m$, the protocol jumps downwards.

\section{Concluding Discussion}

In summary, we have calculated optimal protocols yielding the minimal mean work for underdamped Langevin dynamics in two different model potentials. Surprisingly, these optimal protocols involve jumps and delta peaks at the initial and final times $t_i=0$ and $t_f$. 
While we have shown that the singularities in the optimal protocol appear for harmonic potential, there is no reason to believe that this feature generically vanishes for anharmonic potentials. In fact, in the overdamped limit, a recent study has shown that initial and final jumps are also present in a simple anharmonic potential \cite{then08}.
At first sight, such singularities seem to be unphysical since neither jumps nor delta peaks can be implemented in real experiments. Still our theoretical result is an important insight because it implies that there exists no optimal continuous protocol. Every such protocol could be improved by even steeper gradients mimicking the jumps and delta peaks at the beginning and end. If there was an experimental constraint on the allowed maximum rate of change in $\lambda$, $|\dot \lambda| < r$, the minimal work would still be achieved by a protocol which looks roughly like the optimal one, with the jumps and delta peaks replaced by their best approximation consistent with $|\dot \lambda| < r$ (\textit{e. g.} steep straight lines instead of jumps). Thus, it should be possible to exploit our results for real experiments.

Our results may also be used in steered MD simulations. Even though we here have calculated optimal protocols for underdamped Langevin dynamics, there is no reason to believe that other thermostats frequently used in MD simulations would yield qualitatively different results for the optimal protocol. 
 We have neglected memory effects by assuming white noise. While there are systems for which the underdamped Langevin equation is an appropriate physical description \cite{doua06}, it might still be interesting to see how our results are altered when considering memory effects.

 The optimal protocol for a minimal mean work is not strictly equivalent to a protocol leading to an optimized free energy estimate. However, it has been found that the latter shares the same features (jumps at the boundaries) for overdamped Langevin dynamics \cite{dell08}. Moreover, the optimal protocol leads to improved estimates of the free energy difference in both of our (underdamped) case studies. For case study I, the work distribution is Gaussian \cite{mazo99, tani08} for which it has been shown that the error in the estimate of the free energy difference decreases with decreasing mean work \cite{gore03}. In our second case study, for which the work distribution is no longer Gaussian, the error in the estimate of the free energy difference is indeed lower for the optimal protocol compared to a linear protocol, see Fig. \ref{fig_dF} and its caption for technical details. For thermodynamic integration, it is obvious that a minimum mean work leads to the best estimate of the free energy difference. We thus conjecture that appropriate singularities at the boundaries generically improve free energy calculations from either fast growth methods or thermodynamic integration.

For determining the optimal protocol for an unknown potential, we envisage an adaptive procedure in which trial protocols (including estimated singularities) are successively improved in an iterative fashion guided by the monitored work values. It might also prove beneficial to use the optimal moving  trap protocol (case study I)  rather than a linear protocol in simulations of (protein) pulling experiments.

\appendix
\section{Solution of the optimization problem in case study II}
\label{app1}

In this appendix, we give a detailed analysis of the numerical solution of the optimization problem. In order to minimize the integral in Eq. (\ref{W20}), the constraint [Eq. (\ref{constr})] is included in an effective Lagrangian $\mathcal{L}_{\rm{eff}}\equiv \mathcal{L} - \alpha(t)\mathcal{G}$ through a Lagrange multiplier $\alpha(t)$. Then the Euler-Lagrange equations whose solutions minimize the integral in Eq. (\ref{W20}) are obtained from 
\begin{equation}
\label{EL_gen}
\frac{\partial \mathcal{L}_{\rm{eff}}}{\partial w}+
\frac{d^2}{dt^2}\frac{\partial \mathcal{L}_{\rm{eff}}}{\partial \ddot{w}}=
\frac{d}{dt}\frac{\partial \mathcal{L}_{\rm{eff}}}{\partial \dot{w}}
, \;\;\;\;
\frac{\partial \mathcal{L}_{\rm{eff}}}{\partial z}=\frac{d}{dt}\frac{\partial \mathcal{L}_{\rm{eff}}}{\partial \dot{z}},
\end{equation}
which, together with the constraint $\mathcal{G} = 0$, define a system of three differential equations for $w$, $z$ and $\alpha$.
By defining the useful new variable 
\begin{equation}
\label{def_mu}
\mu \equiv zw-\frac{m^2}{4}\dot{w}^2
\end{equation}
we can write the initially cumbersome differential equations (\ref{EL_gen}) after a tedious manipulation in the following reduced form
\begin{eqnarray}
\ddot{w} & = & \frac{\dot{w}^2}{2w} - \frac{2}{m^2}\frac{\mu}{w} +2Tw\alpha +\frac{2T}{m}  \label{ELv} , \\
\dot{\mu} & = & - \frac{2\gamma}{m}\mu+2\gamma T w , \\
\dot{\alpha} & = & \frac{2\gamma}{m}\alpha+ \frac{1}{m}\frac{\dot{w}}{w^2}.
\end{eqnarray}
These equations have no analytical solution but they can easily be solved numerically for given initial conditions $w(0^+)$, $\dot{w}(0^+)$, $\mu(0^+)$ and $\alpha(0^+)$.
It is important to note that some of these initial conditions are not fixed by the initial equilibrium conditions $w(0) = T/\lambda_i$, $\dot w (0) = 0$, $\ddot w(0) = 0$, $z(0) = m T$, but can be realized by additional discontinuities in the respective quantities at the boundaries. If such discontinuities do not change the value of the integral in Eq. (\ref{W_stiff}), they do not affect the optimization of the integral via the Euler-Lagrange equations and hence the respective initial conditions should (in a first step) be treated as free parameters. Since the Langrangian does not depend on $\ddot w(t)$, discontinuities in $\dot w(t)$ and $\ddot w(t)$ can occur at the boundaries. However, a jump in the mean squared position $w(t)$ would affect the integral in Eq. (\ref{W_stiff}) and thus $w(t)$ must be chosen to be continuous at the boundaries, enforcing  $w(0^+)=w(0)\equiv w_0=T/\lambda_i$. Likewise, discontinuities in $z(t)$ can occur at the boundaries. However, the initial values $z(0^+)$ and $\dot w(0^+)$ are related by the constraint $\mathcal{G}=0$. Integrating this constraint
\begin{equation}
\lim_{\epsilon \to 0} \int_{t-\epsilon}^{t +\epsilon} dt' \mathcal{G}=0
\end{equation}
leads to
\begin{equation} \label{magicrel}
\left[ wz \right]^{t^+}_{t^-}=\frac{m^2}{4}\left[\dot{w}^2\right]^{t^+}_{t^-}.
\end{equation}
When applied at $t=0$
it yields
\begin{equation}\label{magicrel0}
\frac{T}{\lambda_i}[z(0^+)-mT]=\frac{m^2}{4}\dot{w}^2(0^+).
\end{equation}
We consider a (possible) discontinuity through the parameter $s_1$ in
\begin{equation}\label{z0+}
z(0^+)\equiv mT s_1.
\end{equation}
With Eq. (\ref{magicrel0}), the jump in the derivative of $w$ at the initial time as a function of $s_1$ becomes
\begin{equation}
\dot{w}(0^+)=\pm 2T \sqrt{\frac{s_1-1}{m\lambda_i}}.
\end{equation}
In the case in which $\lambda_i<\lambda_f$, the correct sign is the negative one. Note that the last equation implies $s_1>1$, so that at the initial time and given the equilibrium initial distribution, it is not possible to have a decrease in the mean squared momentum. 
From Eqs. (\ref{magicrel0}) and (\ref{def_mu}) we also find  
\begin{equation}
\mu(0^+)= mT^2 /\lambda_i. 
\end{equation}

Secondly, we define a new free parameter $s_2$ in
\begin{equation}
\dot{y}(0^+)\equiv T s_2,
\end{equation}
which, from the evolution equation (\ref{ydot}), directly yields 
$\ddot{w}(0^+)= \frac{2T}{m} s_2$.
Then, writing Eq. (\ref{ELv}) at $t=0^+$ and inserting the above values,
the initial value of the Lagrange multiplier needed to solve the Euler-Lagrange equations is
\begin{equation}
\alpha(0^+)=\frac{\lambda_i}{mT}(s_2-s_1+1).
\end{equation}
Last, from the evolution equations (\ref{wdot})-(\ref{ydot}) we find the relative value of the initial jump in the protocol as a function of $s_1$ and $s_2$:
\begin{equation}
\frac{\Delta \lambda_i}{\lambda_i}\equiv
\frac{\lambda(0^+)-\lambda_i}{\lambda_i}=s_1-s_2-1+ \gamma \sqrt{\frac{s_1-1}{m\lambda_i}}.
\end{equation}

At the end of the process,
the value of $z(t_f)$ is allowed to jump again. 
Recalling Eq. (\ref{magicrel}) applied now at the final time $t=t_f$ and isolating $z(t_f)$,  we obtain
\begin{equation}
z(t_f)=z(t_f^-)+\frac{m^2}{4}\frac{\dot{w}(t_f)^2-\dot{w}(t_f^-)^2}{w(t_f)}.
\end{equation}
Every quantity on the right hand side of the last equation except for $\dot{w}(t_f)$ is fixed by the solution of the Euler-Lagrange equation. The minimum value for $z(t_f)$, which leads to the minimal contribution to the work in Eq. (\ref{W2}), is reached for $\dot{w}(t_f)\equiv s_3=0$.

For a comparison of the present case with its overdamped analogue \cite{schm07}, one can formally integrate the differential equations for $\mu$ and $\alpha$ and plug them into Eq. (\ref{ELv}) to obtain the following integro-differential equation for $w$,
\begin{equation}\label{intdif}
\left( \ddot{w}-\frac{\dot{w}^2}{2w}\right) = \frac{2T} {m} \left [ 
f(t)\mathcal{A}-\frac{\mathcal{B}}{f(t)}+f(t)(1+s_2-s_1) \right ]
\end{equation}
where $f(t)\equiv \frac{w(t)}{w_0}e^{2\gamma t/m}$ and 
\begin{equation}
\mathcal{A}=1-\frac{2\gamma}{m}  \int_0^{t} \frac{1}{f(t')} dt', 
\;\;\;\;\; 
\mathcal{B}=1+\frac{2\gamma}{m} \int_0^{t} f(t') dt'.
\end{equation}
In the overdamped limit, the Euler-Lagrange equation is given by $\ddot{w}-\dot{w}^2 / 2w=0$. Including inertia leads to nonvanishing terms on the right hand side of Eq. (\ref{intdif}). However, taking the corresponding overdamped limit $\tilde m \to 0$ in Eq. (\ref{intdif}) yields the overdamped Euler-Lagrange equation only after optimizing the parameters $s_1$ and $s_2$.

Combining Eqs. (\ref{wdot})-(\ref{ydot}), the work $W$ [Eq. (\ref{W1})] can be written as  
\begin{eqnarray} \label{W2}
W  =
\left[ \frac{\lambda w}{2} +\frac{z}{2m} \right]_{0}^{t_f}  -\frac{\gamma T}{m}t_f + \frac{\gamma}{m^2}\int_0^{t_f}dt z.
\end{eqnarray}
To calculate the integral, we insert the solution of the Euler-Lagrange equations for $z$, which depends on $s_1$ and $s_2$. 
Then, we need to insert the boundary values for $w$ and $z$ at $t=0$ and $t=t_f$.  In a last step, the work is optimized with respect to the free parameters $s_1$ and $s_2$.


\newpage

\subsection*{Figure captions}

\begin{figure*}[h!]
\begin{center}
\includegraphics[angle=0, width=0.45 \linewidth]{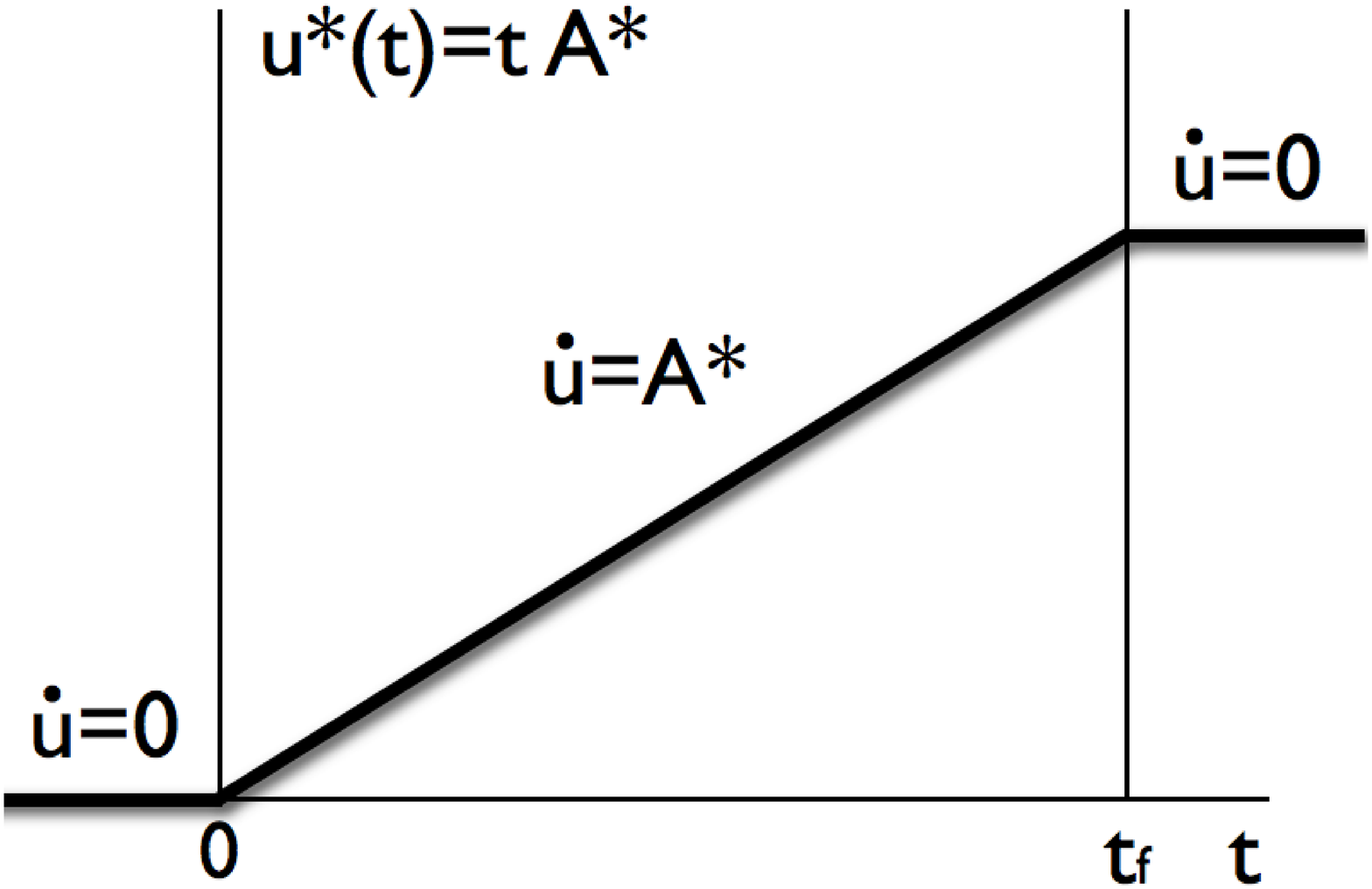}
\space  $\;\;\;\;\;\;\;\;\;\;$ 
\includegraphics[angle=0, width=0.45 \linewidth]{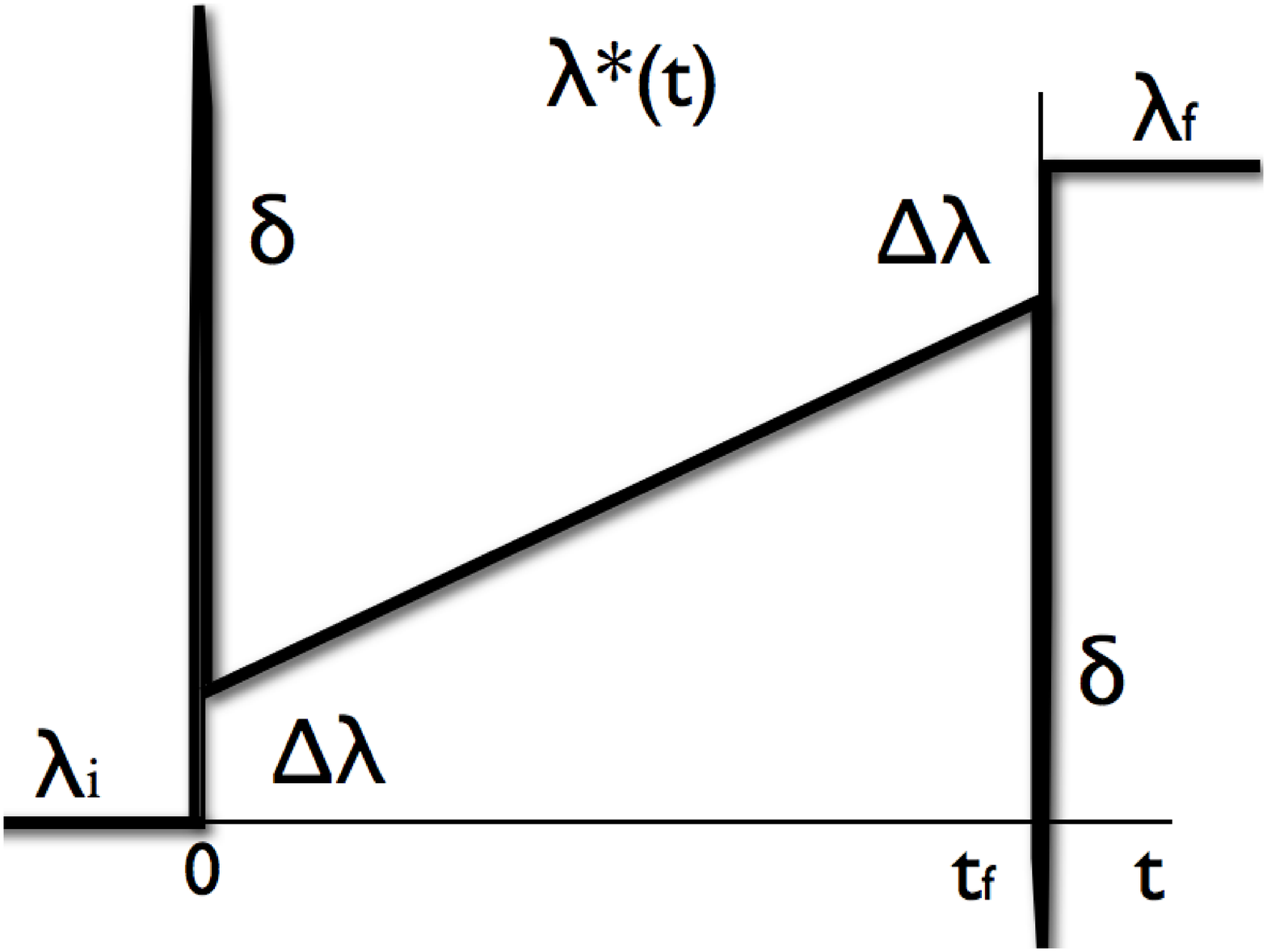}
\caption{} 
\label{fig1}
\end{center}
\end{figure*}

\begin{figure}[h!]
\begin{center}
\includegraphics[angle=0, width= 0.99 \linewidth]{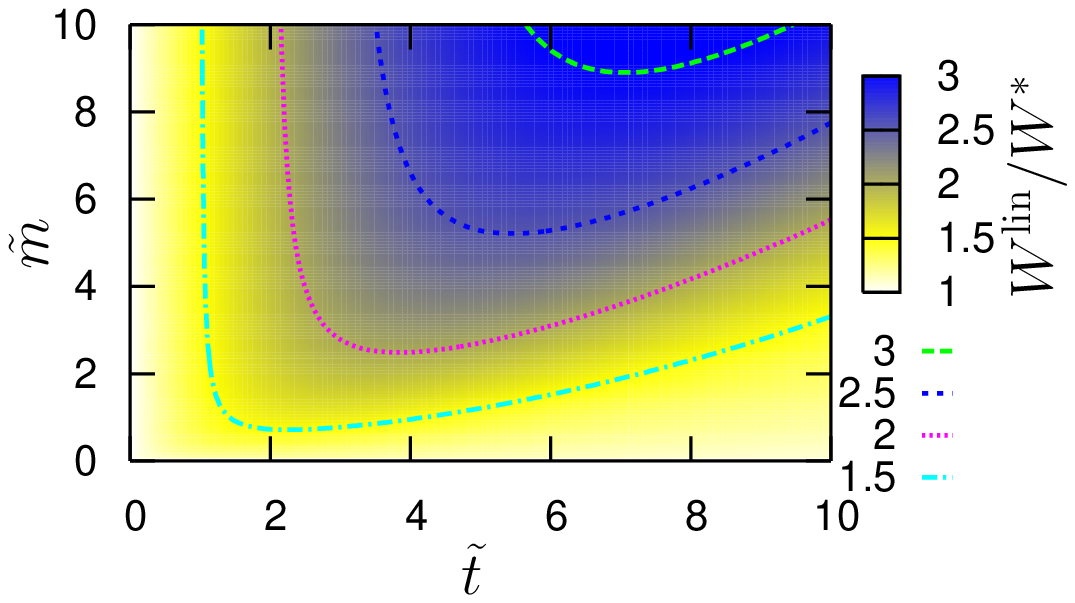}
\caption{} 
\label{fig2}
\end{center}
\end{figure}

\begin{figure*}
\begin{center}
\includegraphics[angle=0, width=0.49 \linewidth]{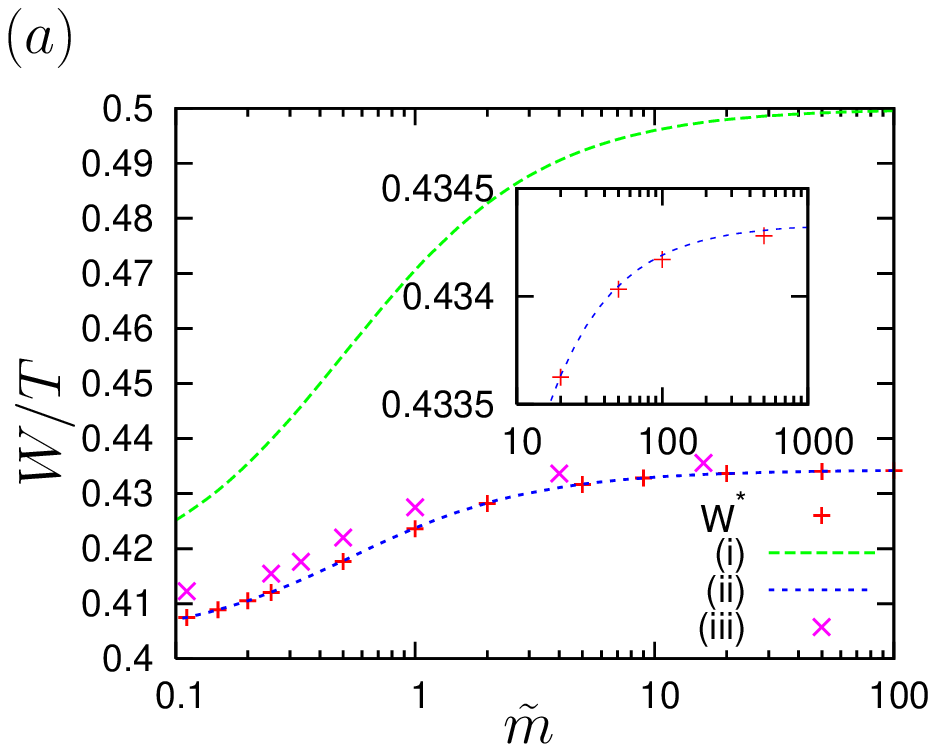}
\includegraphics[width = 0.46 \linewidth] {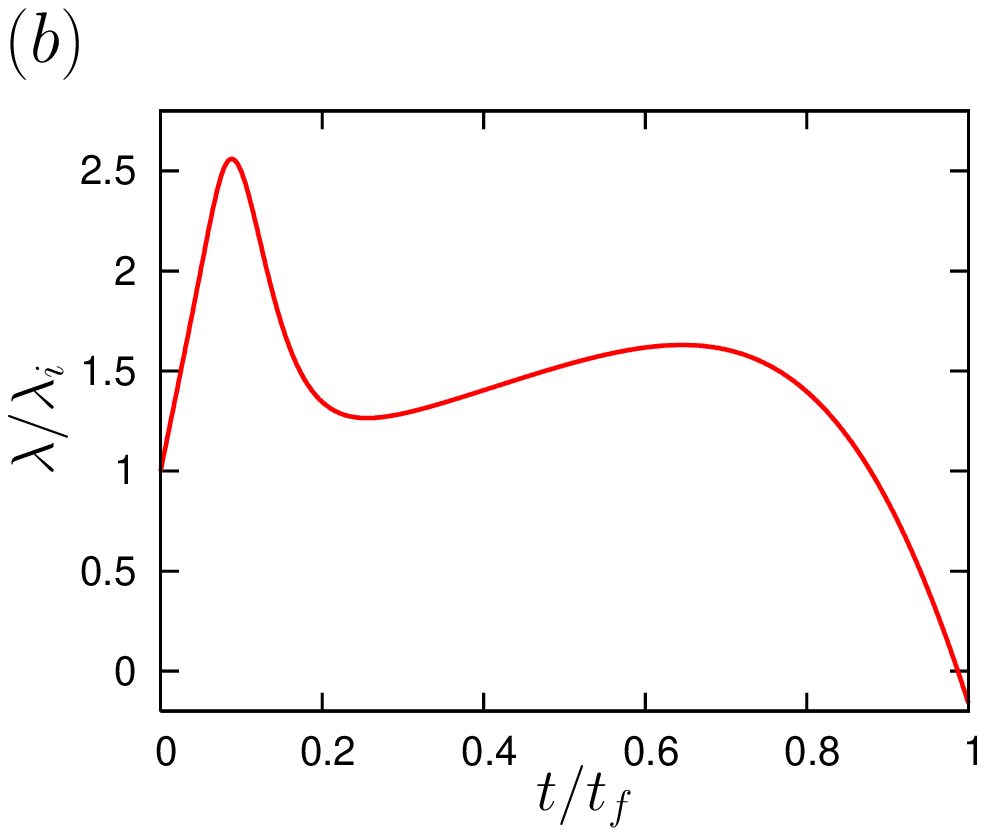}
\includegraphics[angle=0, width=0.49 \linewidth]{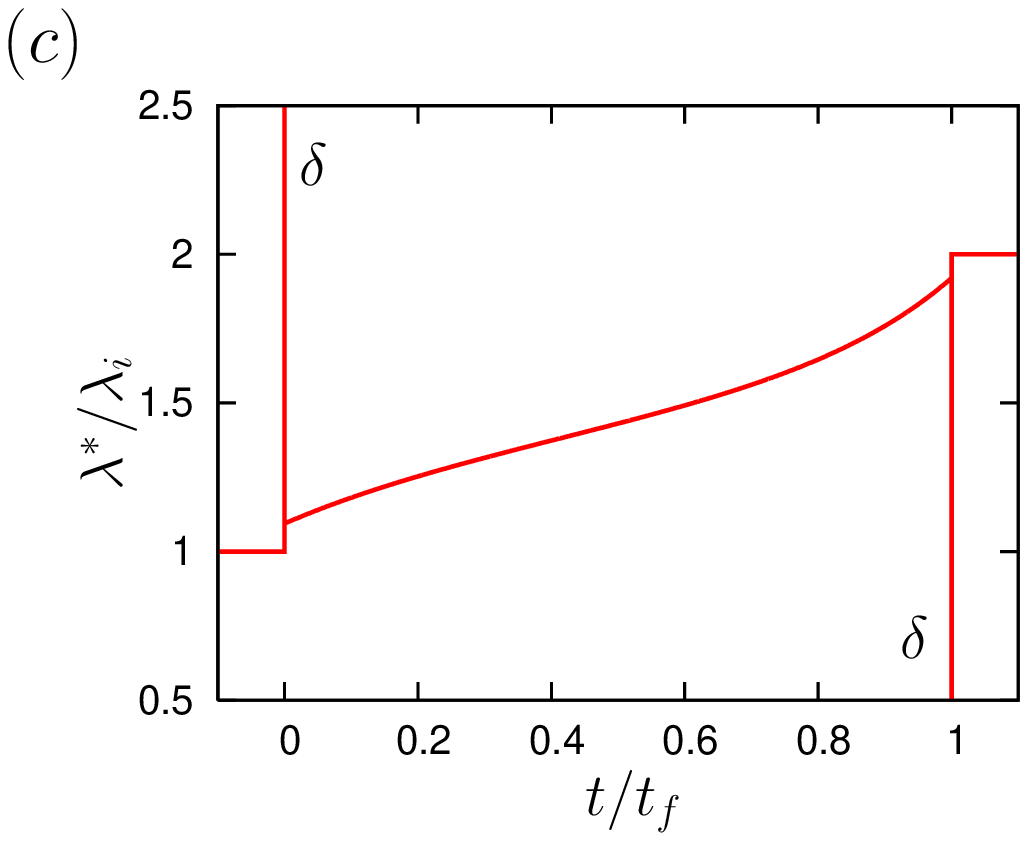}
1\includegraphics[width = 0.49 \linewidth] {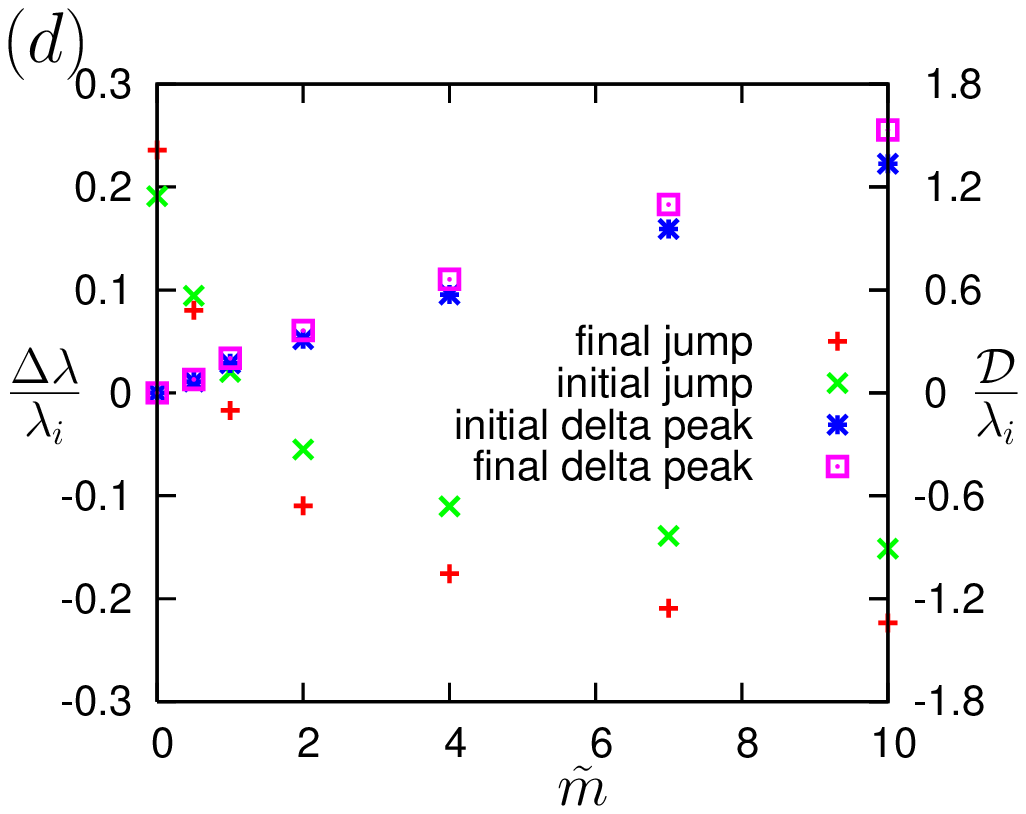}
\caption{} 
\label{lam*}
\end{center}
\end{figure*}

\begin{figure*}
\begin{center}
\includegraphics[width = 0.50 \linewidth] {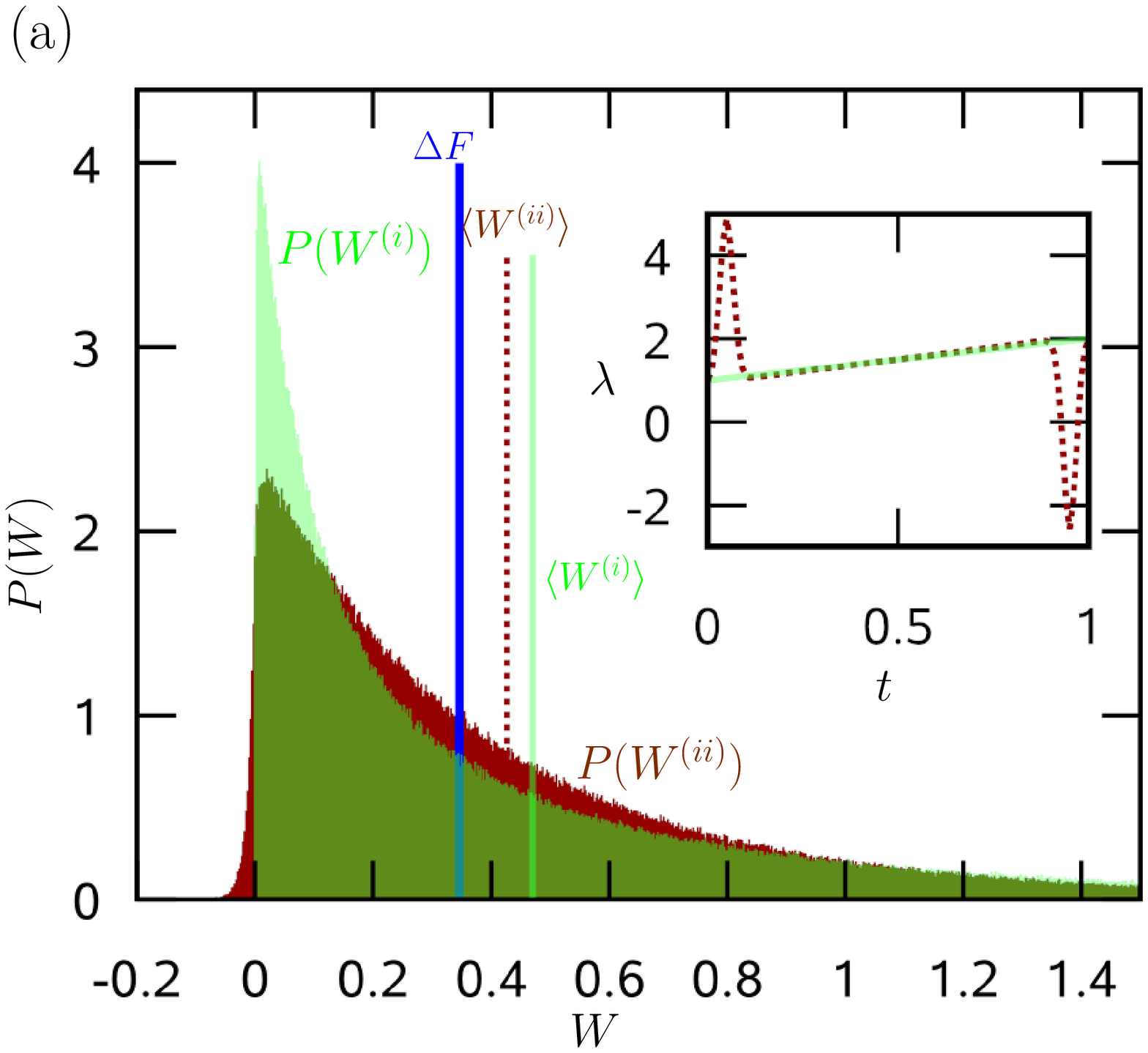}
\includegraphics[angle=0, width=0.49 \linewidth]{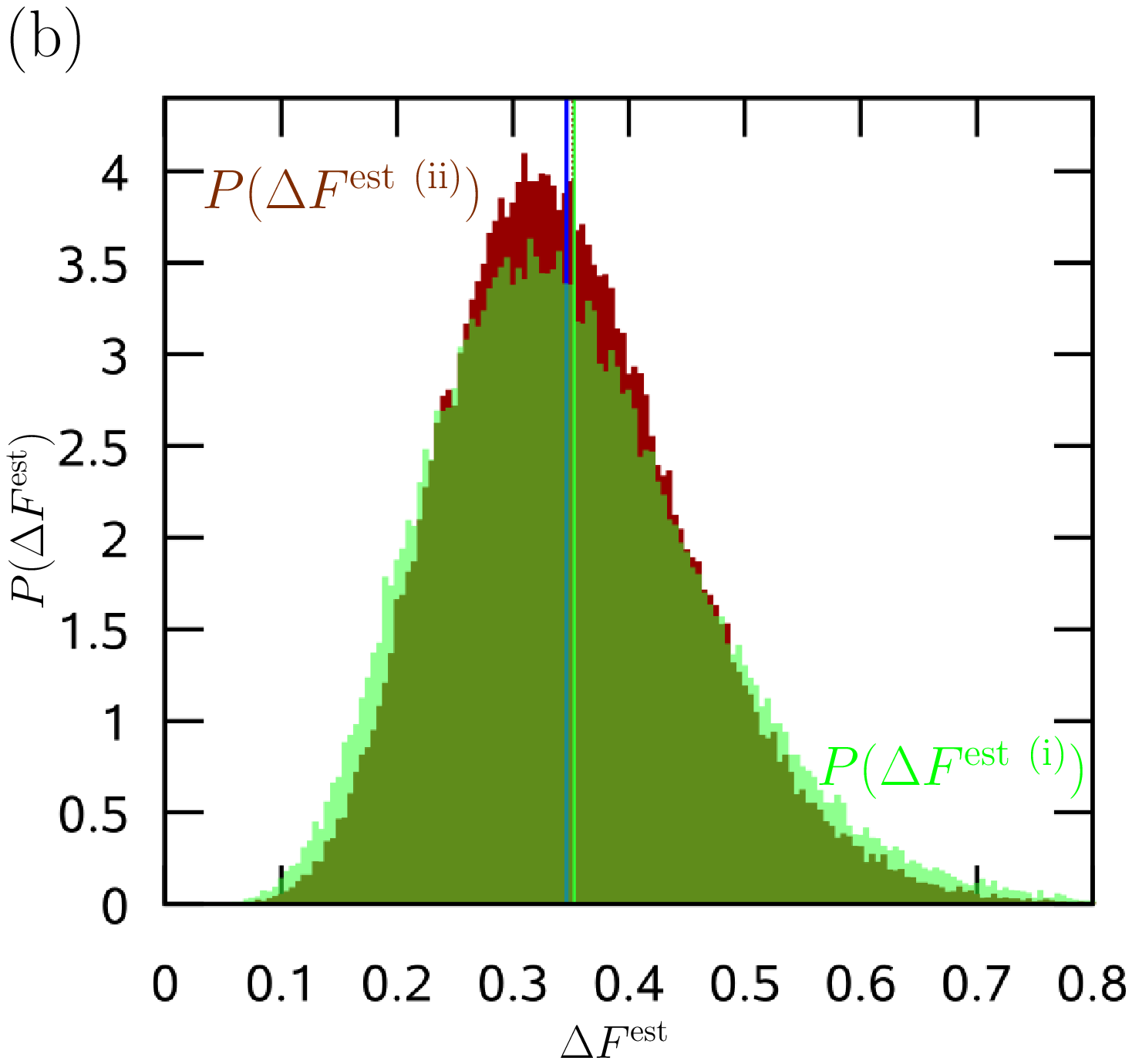}
\caption{} 
\label{fig_dF}
\end{center}
\end{figure*}

Figure \ref{fig1} : Scheme of the optimal mean position $u^*(t)$  and protocol $\lambda^*(t)$.
\vskip 2cm
Figure \ref{fig2} : Ratio between mean work $W^{\rm{lin}}$ spent using the continuous linear protocol $\lambda^{\rm{lin}}(t)$ and optimal work $W^*$ as a function of the dimensionless parameters $\tilde{m}\equiv m k/\gamma^2$ and $\tilde{t} \equiv t_f k/\gamma$.
\vskip 2cm
Figure \ref{lam*} : Optimization results for case study II for $\tilde{t} = 1$ and $\tilde {\lambda} = 2$. (a) Mean work $W^*$ in units of $T$ as a function of the rescaled mass $\tilde{m}$ compared to (i) a linear protocol, (ii) a protocol leading to $w(t)$ given by Eq. (\ref{w_harm}), and (iii) a continuous (except for a final jump) protocol leading to $w(t)$ given by Eq. (\ref{w_poly}) with adjusted parameters to yield a minimal work (see main text for details). (b) Protocol (iii) with optimized parameters for $\tilde m = 1$. (c) Optimal protocol $\lambda^*(t)$ for $\tilde{m} = 0.5$. (d) Jump heights $\Delta \lambda$ and amplitudes $\mathcal{D}$ of delta peaks (in rescaled time $t / t_f$) for the optimal protocol as a function of the rescaled mass $\tilde m$.
\vskip 2cm
Figure \ref{fig_dF} : Comparison of free energy estimates for case study II for a linear protocol and a continous approximation to the optimal protocol for $\tilde t = 1, \tilde m = 1, \tilde \lambda = 2$. The data were obtained from Langevin simulation of $10^6$ trajectories for each protocol with $\gamma = 1, m = 1, T=1, \lambda_i = 1$.  (a) Distribution $P(W)$ of work values $W$ for the two protocols shown in the inset: (i) the linear protocol $\lambda(t) = 1+t$ and (ii) the linear protocol with additional continuously approximated delta singularities. The columns show the free energy difference $\Delta F \simeq 0.3466$ and the mean work values $\mean{W^{(i)}} \simeq 0.4700 (\pm 0.0006)$, $\mean{W^{(ii)}} \simeq 0.4270 (\pm 0.0005)$ which both are consistent with a direct evaluation based on Eq. (\ref{W1}). (b) Histogram of $10^5$ Jarzynski estimates for the free energy difference $\Delta F^{\rm{est}} \equiv - (1 / \beta) \ln \left [\sum_{i=1}^N \exp (-\beta W_i) / N \right ]$ obtained from $N = 10$ single trajectory work values $W_i$ each. The mean squared error (MSE) of these estimates consists of two parts \cite{gore03}: the systematic error (bias) $B = \mean{\Delta F^{\rm{est}}} - \Delta F$ and the statistical error $\sigma = \sqrt{\rm{Var} (\Delta F^{\rm{est}} )}$. The columns show the free energy difference and the mean value of the estimates obtained from the two protocols. Since the bias ($B^{(i)} = 0.0066 (\pm 0.0004)$, $B^{(ii)} = 0.0052 (\pm 0.0003)$) can be neglected for both protocols, the MSE is dominated by the statistical error ($\sigma^{(i)} = 0.119$, $\sigma^{(ii)} = 0.104$) which is smaller for protocol (ii). 

\newpage

\newpage

\newpage

\newpage

\end{document}